\documentclass[sigconf,review=false,nonacm]{acmart}
\AtBeginDocument{%
  }

\usepackage{cleveref}

\newcommand{\cognicrypt}{\emph{CogniCrypt}}

\newcommand{\cambench}{\emph{CamBench\textsubscript{CAP}}}
\newcommand{\cryptoApiBench}{\emph{CryptoAPI-Bench}}

\newcommand{\predictor}{\emph{FP\textsubscript{Predictor}}}

\setcopyright{cc}
\setcctype{by}
\copyrightyear{2026}
\acmYear{2026}

\begin{document}

\title{\predictor\ -- False Positive Prediction for Static Analysis Reports}
\author{Tom Ohlmer}
\email{tohlmer@mail.uni-paderborn.de}
\orcid{0009-0005-0644-7895}
\affiliation{
	\institution{Heinz Nixdorf Institute at Paderborn University}
	\city{Paderborn}
	\state{Nordrhein-Westfalen}
	\country{Germany}
}

\author{Michael Schlichtig}
\email{michael.schlichtig@uni-paderborn.de}
\orcid{0000-0001-6600-6171}
\affiliation{
	\institution{Heinz Nixdorf Institute at Paderborn University}
	\city{Paderborn}
	\state{Nordrhein-Westfalen}
	\country{Germany}
}

\author{Eric Bodden}
\email{eric.bodden@uni-paderborn.de}
\orcid{0000-0003-3470-3647}
\affiliation{
  \institution{Heinz Nixdorf Institute at Paderborn University \& Fraunhofer IEM}
  \city{Paderborn}
  \state{Nordrhein-Westfalen}
  \country{Germany}
}

\begin{abstract}
Static Application Security Testing (SAST) tools play a vital role in modern software development by automatically detecting potential vulnerabilities in source code.
However, their effectiveness is often limited by a high rate of false positives, which wastes developer´s effort and undermines trust in automated analysis.
This work presents a Graph Convolutional Network (GCN) model designed to predict SAST reports as true and false positive.
The model leverages Code Property Graphs (CPGs) constructed from static analysis results to capture both, structural and semantic relationships within code.
Trained on the CamBenchCAP dataset, the model achieved an accuracy of 100\% on the test set using an 80/20 train–test split.
Evaluation on the CryptoAPI-Bench benchmark further demonstrated the model’s practical applicability, reaching an overall accuracy of up to 96.6\%.
A detailed qualitative inspection revealed that many cases marked as misclassifications corresponded to genuine security weaknesses, indicating that the model effectively reflects conservative, security-aware reasoning.
Identified limitations include incomplete control-flow representation due to missing interprocedural connections.
Future work will focus on integrating call graphs, applying graph explainability techniques, and extending training data across multiple SAST tools to improve generalization and interpretability.

\end{abstract}


\maketitle

\noindent\textit{Accepted at STATIC '26 (ICSE 2026 workshop).}

\section{Introduction}
Static Application Security Testing (SAST)~\cite{chess2007secure} tools have become an integral part of software development workflows.
These tools perform static code analyses to automatically detect potential vulnerabilities in source code, helping developers protect sensitive data and maintain the integrity of software systems.
However, as cryptographic and other security-critical APIs become increasingly complex, developers often struggle to use them correctly \cite{AndroidMissuseStudy,JavaCrypStat}.
Empirical studies have shown that cryptographic APIs are misused in up to 85–88\% of examined applications, yet not all of these misuses correspond to exploitable vulnerabilities \cite{AndroidMissuseStudy,JavaCrypStat}.
Generally, developers struggle in using cryptographic APIs securely for various reasons~\cite{nadi2016hoops}.
Chen et al.~~\cite{chen2024towards} highlight that static analysis approaches tend to over-approximate potential flaws, making it difficult to distinguish between false and true positives in vulnerability reports.

This abundance of false positives poses a significant challenge.
Development teams often spend substantial time and resources investigating and dismissing false alarms, which delays the remediation of genuine vulnerabilities, increases overall costs, and reduces trust in automated security tools \cite{nachtigall2022usability}.
Existing SAST tools often rely on rule-based or pattern-matching techniques~\cite{kruger2019crysl}.
While effective for identifying common security flaws, these methods struggle to capture the complex relationships within source code that determine whether a reported issue is actually exploitable \cite{chen2024towards,JavaCrypStat,SASTComparison}.

To address the problem of too many false positives, Tripp et al.~\cite{tripp2008aletheia} suggest training a model to predict whether an analysis report is a false positive.
In this paper we aim to help developers to focus on fixing real vulnerabilities by automatically differentiating between false and true positives.

Following this approach, we contribute \predictor, a trained model to predict warning reports of Java cryptographic API misuses.
Our approach is based on a Graph Convolutional Network (GCN) model~\cite{gnn} that leverages Code Property Graphs (CPGs)~\cite{cpgvulnerability} representing the structural and semantic relationships in code.

Despite their potential, applying graph-based machine learning for the classification of SAST findings remains relatively underexplored.
We use \cognicrypt~\cite{kruger2019crysl} as a representative SAST tool to provide the analysis results as input for \predictor.
\predictor\ is trained and tested using \cambench~\cite{schlichtig2022cambench}, which consists of secure and vulnerable files as well as the ground truth.
We evaluate \predictor\ using a different benchmark, \cryptoApiBench~\cite{cryotiapibench}, also comprised of secure and vulnerable files, and ground truth, to assess its generalization to other cryptographic misuse scenarios.
By learning from the structural and contextual information encoded in CPGs, the proposed method aims to improve the accuracy of false positive detection and enhance the efficiency of static analysis workflows.
Overall, \predictor\ in its current form demonstrates a promissing approach that we plan to further explore and employ to sort and filter static analysis results.

\label{sec:intro}
\section{Related Work}
Recent advances in machine learning for software security have leveraged  Graph Neural Networks (GNNs) \cite{gnn} to model source code structure.
CPGs combine multiple program views, namely Abstract Syntax Trees (ASTs), Control Flow Graphs (CFGs), and Program Dependence Graphs (PDGs) into a unified representation~\cite{cpgvulnerability}.
CPGs enable learning models to capture both syntactic and semantic dependencies that are essential for vulnerability detection.
Several works have proposed GNN variants, such as Gated Graph Neural Networks (GGNNs) and heterogeneous GNNs, to improve representation power across diverse edge types and relationships~\cite{luo2025detecting}.
Luo et al. introduced a Heterogeneous Attention GNN (HAGNN) and an inter-procedural abstract graph (IPAG) representation to better model relationships across functions and modules.
Their results demonstrated high predictive accuracy for vulnerability detection across large C and Java datasets.
Prior research has focused primarily on predicting vulnerabilities directly from code. 
In contrast to previous GNN-based vulnerability prediction approaches, this work focuses on prediction in SAST reports.
This enables, e.g., to sort or filter SAST reports and thereby help developers to focus on likely true positives.

\label{sec:related}
\section{False Positive Machine Learning Model}
\label{sec:model}
To predict false positives in static analysis reports, we propose a training GCN–based model~\cite{gnn} that classifies the CPG~\cite{cpgvulnerability} of a reported vulnerability as being a true or false positive.
The model leverages CPGs to capture both the structural and semantic characteristics of the code surrounding a reported issue.
Model training takes about five minutes on a PC (Intel 11700k, Nvidia RTX 4090, 64 GB memory).
\begin{figure}[h!]
    \centering
    \includegraphics[width=0.45\textwidth]{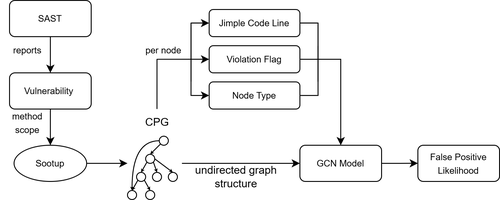}
    \caption{\predictor\ -- pipeline for predicting whether a reported vulnerability is a false positive.}
    \label{fig:FPpipeline}
\end{figure}
\vspace{-1em}

\subsection{Graph Construction}
For each reported vulnerability, a CPG~\cite{cpgvulnerability} is constructed using the static analysis framework SootUp~\cite{karakaya2024sootup} for the corresponding method in which the violation occurs (cf.~\Cref{fig:FPpipeline}).
The CPG represents source code as a unified, graph-based intermediate representation that combines multiple program analysis views, integrating three fundamental graph structures: The AST, the CFG, and the PDG~\cite{cpgvulnerability}.

The AST captures the syntactic structure of the code by representing statements, expressions, and declarations as hierarchical nodes.
It provides the structural backbone of the program and preserves the original code layout and nesting.
The CFG encodes the control-flow relationships between statements, modeling possible execution paths and branching conditions.
Finally, the PDG describes data and control dependencies between program elements, linking statements that influence one another through variable usage or control predicates \cite{cpgvulnerability}.

By merging these representations into a single graph, the CPG serves as input for the GCN-based~\cite{gnn} model and enables joint reasoning over syntactic, control, and data dependencies.
Each node and edge carries semantic information that helps the GCN learn relationships for distinguishing secure and vulnerable code.

The process of constructing a CPG is based on SootUp~\cite{karakaya2024sootup}, which parses the Java source code into the intermediate representation Jimple and subsequently generates the associated CPG.
Within each graph, the node whose Jimple statement corresponds to the reported violation (ground truth or report) is marked as the violation node, serving as a focal point for feature extraction and learning.

\subsection{Feature Representation}
Each node in the CPG~\cite{cpgvulnerability} is enriched with three feature types, capturing both syntactic and structural information:

\begin{itemize}
    \item \textbf{Vectorized Jimple Code} — The raw Jimple statement of each node is encoded using a pre-trained Word2Vec~\cite{mikolov2013efficient} model to capture syntactic and contextual relationships.

    \item \textbf{Node Type Encoding} — The node type (e.g., assignment, call, control flow) is represented using a one-hot encoding scheme to preserve structural information.

    \item \textbf{Violation Flag} — A binary indicator marks whether the node corresponds to the ground truth or reported violation.
\end{itemize}

These features, combined with the inherent structural information of the graph, form the complete input to the GCN~\cite{gnn} model.

\subsection{Model Training}

The GCN~\cite{gnn} processes the CPGs~\cite{cpgvulnerability} by aggregating information from neighboring nodes and learning hierarchical representations of code semantics.
A sigmoid output layer~\cite{gnn} produces a score in $[0,1]$ representing the model’s confidence that a vulnerability report is a false positive.
For demonstration purposes, a threshold of 0.8 is applied, with scores above it classified as false positives and lower scores as true positives.

The model is trained on the benchmark \cambench~\cite{schlichtig2022cambench}, which contains labeled (ground truth with vulnerable line of code) examples of true and false positive vulnerabilities.
An 80/20 train–test split is used to evaluate model performance.
During evaluation, the model achieved an accuracy of 100\% on the \cambench\ test set, demonstrating \emph{a)} the structured approach of how \cambench\ was synthetically generated to contain variations of common vulnerabilities, \emph{b)} the model's ability to generalize, and \emph{c)} the model's ability to predict false positives from genuine vulnerabilities.

\section{Experimental Setup}
To evaluate the performance and generalization of \predictor\ beyond the training data, we apply it to \cryptoApiBench~\cite{cryotiapibench}, a benchmark for assessing SAST tools on cryptographic API misuses.
Each case is labeled with a vulnerability type and a Boolean flag indicating whether it represents a true vulnerability or a benign instance.

The benchmark contains both intra- and inter-class vulnerability scenarios. 
As \predictor\ was trained exclusively on single-class vulnerabilities~\cite{schlichtig2022cambench}, only intra-class cases are evaluated, excluding the 21 inter-class cases in \cryptoApiBench.

For testing, all benchmark cases are analyzed using \cognicrypt~\cite{kruger2019crysl} in version 5.0.2 as a representative static analysis tool.
The resulting reports are compared against the ground-truth labels from \cryptoApiBench\ to classify reports as either true positives or false positives, according to the following criteria:

\begin{itemize}
\item \textbf{True Positive (\emph{TP}):} \cognicrypt\ reports a vulnerability, and \cryptoApiBench\ marks the case as a true vulnerability.
\item \textbf{False Positive (\emph{FP}):} \cognicrypt\ reports a vulnerability, but \cryptoApiBench\ marks the case as secure.
\end{itemize}

Each reported case is then transformed into a CPG~\cite{cpgvulnerability}, where the reported violation line from \cognicrypt\emph{'s} analysis is mapped to the corresponding node in the graph.
These CPGs serve as input to \predictor.
The predictions are subsequently compared to the \emph{} ground truth labels to determine whether it correctly identifies the nature of each report.
Based on this comparison, the accuracy of \predictor\ on the \cryptoApiBench\ is computed as follows:

\begin{equation}
\label{eq:accuracy}
\text{Accuracy} = 
\frac{TP_{\text{correct}} + FP_{\text{correct}}}
{TP_{\text{correct}} + FP_{\text{correct}} + TP_{\text{incorrect}} + FP_{\text{incorrect}}}
\end{equation}\vspace{0.2em}

Here, \(TP\) and \(FP\) denote the sets of ground truth \textit{true positive} and \textit{false positive} cases, respectively.  
The subscripts \(\text{correct}\) and \(\text{incorrect}\) indicate whether a case was classified correctly or incorrectly by the model with respect to the ground truth.

\label{sec:expsetup}
\section{Results and Discussion}
We present our result in \Cref{tab:results} and provide an artifact~\cite{artifact} comprising of \predictor, the experiment pipeline, and results including the manual inspection.

\begin{table}[h]
	\centering
	\begin{tabular}{l|c|c|c}
		\hline
		Ground Truth & CC report & \(FP\) predictions & Inspection \\
		\toprule
		Vulnerable (TP) & 91 & 2 of 91 FP & 89 + 2 \\ 
		Secure (FP) & 27 & 1 of 27 FP & 1 + 22 \\ 
		\bottomrule
	\end{tabular}
	\caption{Experiment result. Ground truth for \cryptoApiBench~\cite{afrose2022evaluation} -- a CC (\cognicrypt~\cite{kruger2019crysl}) report for a vulnerable file is a \emph{TP} and for a secure file a \emph{FP}. \emph{FP} predictions denotes the number of CC reports predicted to be \emph{FPs}. Inspection denotes the number of cases that were correctly predicted after manual inspection, despite the result differing from the ground truth.} 
	\label{tab:results}
\end{table}
\vspace{-2em}
At first glance, the model’s performance on the \cryptoApiBench~\cite{cryotiapibench} dataset appeared unconvincing when evaluated purely on automated predictions.
Out of 27 false positive reports from \cognicrypt, \predictor\ predicted only one as a false positive, corresponding to an accuracy of approximately 3.7\%.
In contrast, for the true positive cases, the model performed considerably better, correctly identifying 89 out of 91 cases as true positives and misclassifying only two as false positives, resulting in an accuracy of about 97.8\%.

Given the weak performance on the false positive subset, a detailed manual inspection to analyze potential causes for misclassifications was conducted.
A recurring pattern emerged across four test cases that contained conditional (\emph{''if''}) statements or other control-flow dependencies.
These cases appeared to be particularly challenging for the model.
This behavior may stem from an insufficient variety of similar code structures in the \cambench~\cite{schlichtig2022cambench} training dataset or from limitations in the CPG representation itself. %

Further manual examination revealed that several cases initially marked as incorrectly predicted, in fact, represent genuine cryptographic misuses that were mislabeled in the benchmark.
In total, 22 out of 26 such cases exhibited security-relevant patterns that justify the model’s decision to treat them as true positives.
The manual inspection reasoning for all cases is detailed in the artifact~\cite{artifact}.

One representative example is the test case \emph{CredentialInStringCorrected.java}.
Although labeled as non-vulnerable in \textit{CryptoAPI-Bench}, manual inspection revealed that it contains multiple insecure coding patterns.
Specifically, an unprotected key is used in line 17, and in lines 21 and 22, the AES cipher is initialized in CBC mode, which is discouraged due to its susceptibility to padding oracle attacks~\cite{owasp_crypto}.
According to the OWASP Cryptographic Storage Cheat Sheet, AES-CBC mode should be avoided unless combined with additional integrity protection, such as HMAC or GCM authentication.
This is also reflected in \cambench.
Thus, \predictor\emph{'s} prediction reflects a valid interpretation of the code’s security risk.

For three incorrectly predicted cases about predictable seeds \predictor\emph{'s} prediction may also be interpreted as reasonable, as they represent bad practice.
For example, in \emph{PredictableSeedsABPSCase1.java}, \predictor\ produced a sigmoid output of 0.833, surpassing the decision threshold of 0.8 as a false positive.
A manual review showed that the \texttt{SecureRandom} class is used with a hard-coded seed (line 13).
Although this behavior does not necessarily weaken the randomness in practice, it is considered a poor cryptographic practice because it introduces determinism into random number generation.
The official Java documentation~\cite{secure_random_doc} clarifies that invoking \texttt{SecureRandom.setSeed(long)} augments, rather than replaces, the existing seed.
The same holds true for the incorrectly predicted true positive cases: 
\emph{PredictableSeedsABHCase2.java} (violation at line 19) and \emph{PredictableSeedsBBCase1.java} (violation at line 10) which involve the use of non-random seed values with \texttt{SecureRandom}.
\predictor\ produced high sigmoid outputs of 0.9756 and 0.9764, indicating strong confidence. %
However, the fact that these values are not exactly 1.0 suggests that the model has not frequently encountered this type of pattern during training and thus exhibits a degree of uncertainty in its decision.

After reevaluation, up to 22 out of 26 initially misclassified cases were found to contain potential vulnerabilities or poor cryptographic practices that could justify the model’s predictions.
As a result, the effective accuracy on the false positive subset increases from 1 out of 27 (3.7\%) to approximately 23 out of 27 (85.2\%).

Based on the manual inspection, \predictor\emph{'s} overall accuracy in predicting true and false positives increases to 96.6\%.
In total, we observed three cases of bad practices in our manual inspection.
If we do not adapt the ground truth label for these cases, as they are debatable, \predictor\emph{'s} accuracy is 94.1\%.
Further, our model’s predictions align with the ground truth revisions of \cryptoApiBench~\cite{cryotiapibench} by Adriano Torres et al. \cite{cryptoapirevision}.

\label{sec:resultdiscussion}
\vspace{-1em}
\section{Limitations and Threats to Validity}
\label{sec:limitations}

\predictor is trained on the \cambench~\cite{schlichtig2022cambench} dataset, which, although well-structured, may not capture the full diversity of real-world vulnerabilities.
Consequently, the model’s generalizability beyond this dataset is uncertain.
The limited size of 431 labeled cases used for training, 80/20 split for training and testing, further restricts the robustness of the results.
The observed up to 100\% in testing may limit generalizability. 
We try to mitigate this by using a different benchmark, \cryptoApiBench~\cite{cryotiapibench} for evaluation.
However, in manual inspection we find that its ground truth is outdated and contains bad practices, e.g., in the \texttt{PredictableSeeds} cases, \texttt{SecureRandom} remains cryptographically secure if properly initialized, since the static seed is only mixed into the entropy.

Additionally, the current CPG~\cite{cpgvulnerability} representation only models single-class vulnerabilities.
Interprocedural dependencies across multiple classes or methods are not captured, which may limit the model’s ability to reason about more complex, cross-context vulnerabilities
Integrating call graphs could mitigate this limitation in future work.

\section{Conclusion}
In this paper we contribute \predictor, a GCN-based model~\cite{gnn} to predict whether reports of static analysis tools for cryptography are false positives.
After manual inspection, the evaluation shows that \predictor\ effectively learns to adapt to the conservative approach in \cambench\emph{'s} ground truth.
Although \predictor\ initially appeared to perform poorly on the false positive subset of  \cryptoApiBench, a detailed manual reevaluation revealed that many of the supposed incorrect predictions were in fact aligned with valid security reasoning.
By identifying patterns that the benchmark labeled as non-vulnerable but which nonetheless exhibited insecure or questionable cryptographic practices, the model showed an ability to generalize the more conservative ground truth of \cambench.  
Applying this reevaluation, the overall accuracy of \predictor\ increased substantially, reflecting a better alignment between model predictions and realistic security assessments.

Overall, \predictor\ demonstrates that a CGN-based model leveraging CPGs as input can be used to predict false positives in static analysis tool reports.
Further improvements and deeper evaluations are needed to apply this approach to existing static analysis tools.
\predictor\ seems to be a promising foundation to integrate sorting or filtering as a post-processing step to analysis reports to tools such as \cognicrypt~\cite{kruger2019crysl}.

\section{Future Work}
Future work could extend the current approach by incorporating interprocedural control flow into CPGs~\cite{cpgvulnerability}, for example via call graphs, to better capture dependencies across methods and classes.

Graph explanation methods may be used to identify which nodes and substructures drive model predictions and to assess whether these decisions align with established security practices.
Code reduction techniques could further reduce graph complexity by eliminating program parts irrelevant to vulnerability behavior.
Finally, the generalizability of \predictor\ should be evaluated using additional benchmarks, multiple SAST tools, and alternative target domains.

\label{sec:conclusion}

\begin{acks}
	Funded by the Deutsche Forschungsgemeinschaft (DFG, German Research Foundation) – SFB 1119 – 236615297
\end{acks}

\bibliographystyle{ACM-Reference-Format}
\vspace{-1em}
\bibliography{references}
\end{document}